\documentclass{raa}            
\usepackage{enumerate}
\usepackage{graphicx,times}             
\usepackage{graphicx}
\usepackage{natbib}
\usepackage{amssymb,amsmath}
\usepackage{diagbox}
\usepackage{multirow}
\usepackage{soul}
\bibpunct{(}{)}{;}{a}{}{,}

\usepackage[pagebackref=true]{hyperref}
\hypersetup{colorlinks = true, linkcolor = green, anchorcolor = red, citecolor = blue, filecolor = red, pagecolor = red, urlcolor = red}

\begin{document}

   \title{Solar Image Deconvolution by Generative Adversarial Network
}

   \volnopage{Vol.0 (20xx) No.0, 000--000}      
   \setcounter{page}{1}          

   \author{Long Xu
      \inst{1}
   \and Wenqing Sun
      \inst{1}
   \and Yihua Yan
      \inst{1}
   \and Weiqiang Zhang
      \inst{2}
   }

   \institute{Key Laboratory of Solar Activity, National Astronomical Observatories, Chinese Academy of Sciences,
             Beijing 100101, China; {\it lxu@nao.cas.cn}\\
        \and
             College of mathematics and statistics, Shenzhen University, Shenzhen, 518060, China\\
\vs\no
   {\small Received~~20xx month day; accepted~~20xx~~month day}}

\abstract{With Aperture synthesis (AS) technique, a number of small antennas can assemble to form a large telescope which spatial resolution is determined by the distance of two farthest antennas instead of the diameter of a single-dish antenna. Different from direct imaging system, an AS telescope captures the Fourier coefficients of a spatial object, and then implement inverse Fourier transform to reconstruct the spatial image. Due to the limited number of antennas, the Fourier coefficients are extremely sparse in practice, resulting in a very blurry image. To remove/reduce blur, ``CLEAN'' deconvolution was widely used in the literature. However, it was initially designed for point source. For extended source, like the sun, its efficiency is unsatisfied. In this study, a deep neural network, referring to Generative Adversarial Network (GAN), is proposed for solar image deconvolution. The experimental results demonstrate that the proposed model is markedly better than traditional CLEAN on solar images.
\keywords{Deep learning (DL); generative adversarial network (GAN); solar radio astronomy; image reconstruction; aperture synthesis}
}

   \authorrunning{L. Xu, W. Sun, Y. Yan, \& W. Zhang }            
   \titlerunning{Deep learning for image deconvolution}  

   \maketitle

%
%
\section{Introduction}           
\label{sect:intro}

For a single-dish antenna, the spatial resolution is limited by the diameter of the dish, subject to $\lambda/D$, where $\lambda$ represents wavelength and $D$ is the dish diameter. It is a big challenge to construct a large single-dish antenna, considering building materials, building technology, architecture and cost. Aperture synthesis (AS) synthesizes a bunch of small antennas to form a big one. Its spatial resolution is determined by the distance of two farthest antennas, namely maximum baseline, still subject to $\lambda/D$, where $D$ is the maximum baseline. Nowadays, AS has been developed intensively in radio astronomy. Many huge radio telescopes, like world-wide low frequency array (LOFAR), Atacama large millimeter array (ALMA) and square meter array (SKA), domestic MingantU SpEctral Radioheliograph (MUSER) have been constructed. MUSER is a solar dedicated AS telescope with the maximum baseline of 3 km, consisting of 100 small antennas. Each pair of antennas compose of an interferometer, recording a Fourier component at each time. We can have $n\times (n-1)/2$ interferometers given $n$ antennas, thus $n\times (n-1)/2$ Fourier components can be obtained. Taking advantage of the earth rotation, one can get more Fourier components. Nevertheless, the Fourier components are very sparse in practice due to the limited number of antennas, resulting in blurry image.
For an AS telescope, image quality degradation is caused by sparse Fourier sampling in frequency domain. The Fourier sampling is described to a frequency-domain image multiplied by a sampling function as shown in Fig. 1 (a). It is corresponding to a spatial image convolved by a point spread function (PSF) or dirty beam as shown in Fig. 1 (b). The sampling function and the PSF are the Fourier transform pairs. Convolving a clear image with the PSF/dirty beam would result in a dirty image which looks blurry. This is because the PSF has strong sidelobe which would cause signal aliasing. To eliminate aliasing, deconvolution, which is the inverse process of convolution, was employed. In radio astronomy, a category of deconvolution algorithms, namely CLEAN \cite{hogbom1974aperture,wakker1988multi,cornwell2008multiscale}, has been extensively studied.

\begin{figure}
   \centering
   \begin{minipage}{0.4\textwidth}
   \centerline{\includegraphics[width=\textwidth, angle=0]{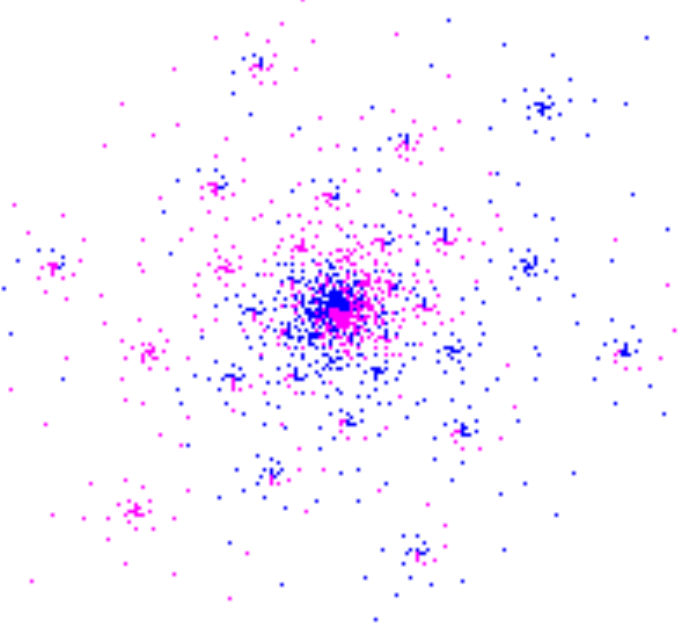}}\centerline{(a) Sampling function}
   \end{minipage}
   \hfill
   \begin{minipage}{0.45\textwidth}
   \centerline{\includegraphics[width=\textwidth,height=0.8\textwidth, angle=0]{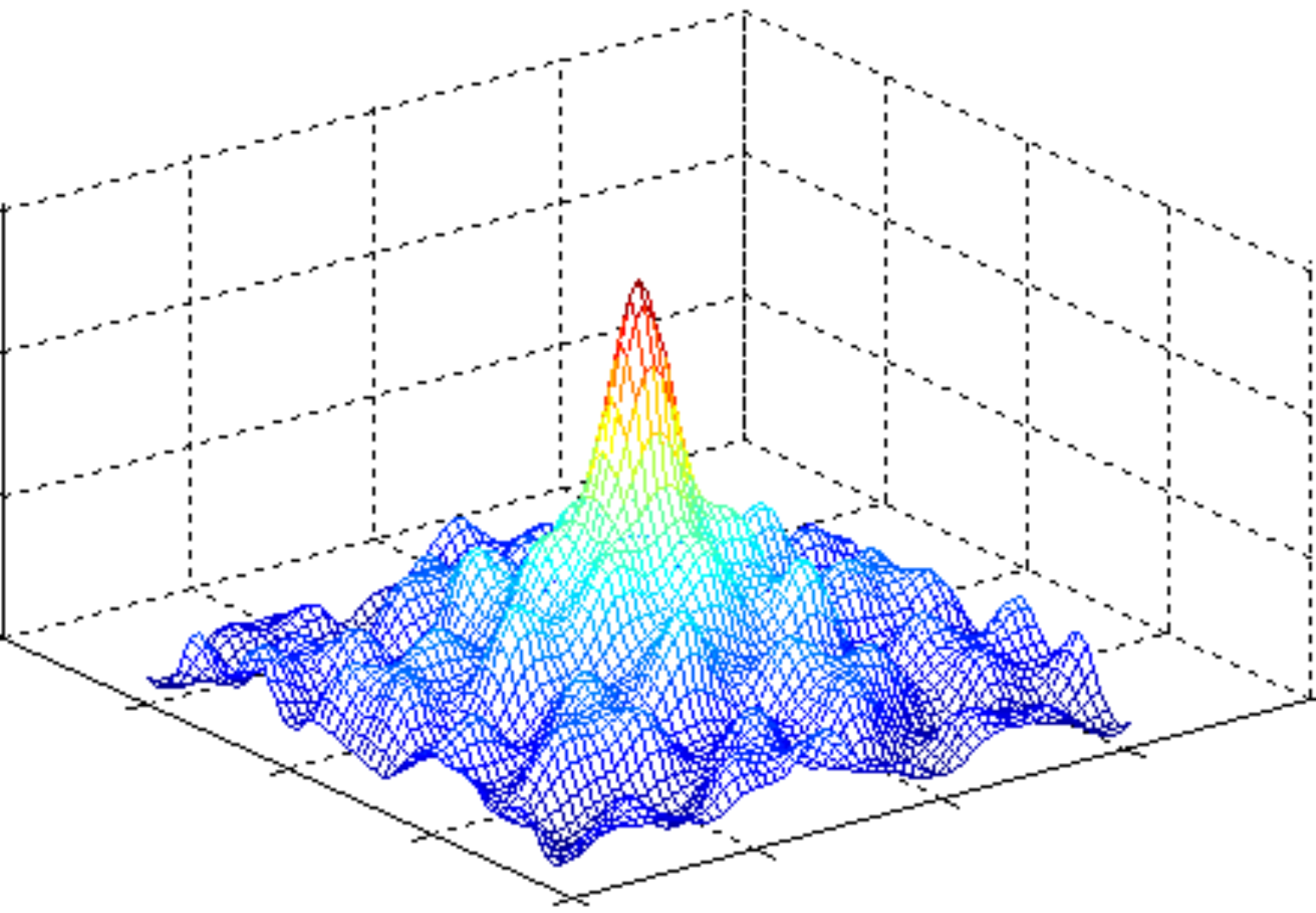}}\centerline{(b) Dirty beam(Point Spread Function)}
   \end{minipage}
   \caption{Sampling function of an Aperture synthesis. }
   \label{fig:1}
\end{figure}

Deconvolution is a deblurring problem in principle. Basically, four categories of image deblurring methods are in the literature. The first one, CLEAN \cite{hogbom1974aperture,wakker1988multi,cornwell2008multiscale}, is mostly used in deconvolution of point source. The second one solves a inverse problem by imposing regularized constraint, such as Total Variation (TV) \cite{ma2008efficient,wen2011primal,rudin1992nonlinear,beck2009fast}, sparseness \cite{elad2006image,zhang2014group,wenger2010sparseri,xu2018image}. The third one is developed on multi-scale signal decomposition \cite{wakker1988multi,cornwell2008multiscale}, such as wavelet, exploiting the multi-scale feature and spectral representation of signal. The last one is learning based method \cite{xiang2015image,su2002hybrid,rubinstein2012analysis,rubinstein2009double,xu2014deep}, which learns signal representation by using machine learning.

The reconstructed image from an AS system usually looks very blurry since highly sparse sampling in Fourier domain. This situation is very common in radio astronomy observation. To address this problem, CLEAN algorithm was widely used. This paper introduces a novel deconvolution algorithm based on Generative Adversarial Network (GAN) \cite{goodfellow2014generative}, to accomplish image deconvolution. The rest of this paper is organized as follows. Section 2 gives the principle of AS. Section 3 gives the details of the proposed deep neural network for image deconvolution. Experimental results are provided in Section 4. And final section draws conclusions.

\section{Aperture synthesis principle}
\label{sect:as}

Given original spatial image by $I(x,y)$, the corresponding image in frequency domain by $V(u,v)$£¬they are the Fourier transform pairs, named by brightness function and visibility function respectively. If there are all Fourier coefficients, $I(l,m)$ can be completely reconstructed. However, the real situation is that $V(u, v)$ is sparsely sampled in Fourier domain. So, a sampled visibility function $V^D(u,v)$ is only available in an AS system, which is represented by
\begin{equation}
V^D(u,v)= V(u,v)\times S(u,v),
\label{eq:1}
\end{equation}
where $S(u,v)$ is the sampling function in frequency domain. Applying inverse Fourier transform to both sides of (1), we can get
\begin{equation}
I^D(l,m)=\iint_\sum V(u,v)S(u,v)\exp(-i2\pi(ul+vm))dudv
\label{eq:2}
\end{equation}
where $I^D(l,m)$ is a dirty image deduced from the Fourier transform of $V^D(u,v)$. Since convolution operation in frequency domain is equivalent to multiplication in spatial domain, (2) can be rewritten into
\begin{equation}
I^D(l,m)=I(l,m)\otimes B^D(l,m),
\label{eq:3}
\end{equation}
where the symbol ``$\otimes$'' denotes convolution operator, and
\begin{equation}
B^D(l,m)=\iint_\sum S(u,v)\exp(-i2\pi(ul+vm))dudv,
\label{eq:4}
\end{equation}
which is the dirty beam or PSF. For easy understanding, we draw a sketch map in Fig. 2 for illustrating the imaging process of an AS system, from both frequency and spatial domain.
\begin{figure}
   \centering
   \includegraphics[width=0.8\textwidth, angle=0]{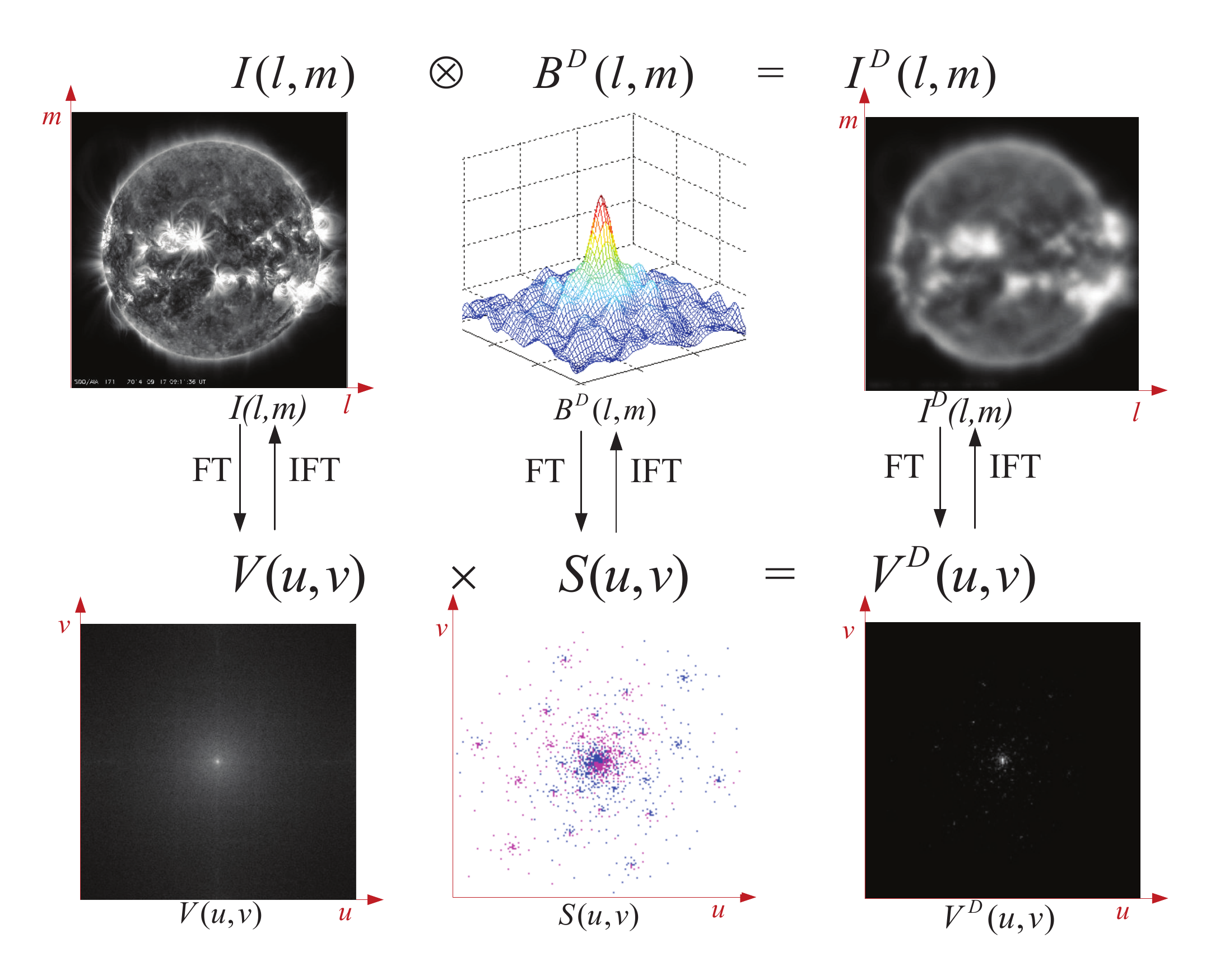}
   \caption{Imaging principle of Aperture synthesis. }
   \label{fig:2}
\end{figure}

From (3), it is only possible to derive $I^D(x,y)$ instead of $I(x,y)$, i.e., dirty image in spatial domain, while the idea image $I(x,y)$ is unavailable since it is polluted by dirty beam $B(x,y)$. To restore $I(x,y)$, we have to delete dirty beam $B(x,y)$ from the left side of (3). This process is usually named ``CLEAN'' deconvolution. For point source, like stellar object, H\"{o}gbom, et. al. proposed a classical H\"{o}gbom CLEAN algorithm, which was witnessed to perform well given dirty beam. However, it was unsatisfied for extended source, like the Sun, so a bunch of algorithms were proposed later, such as multi-resolution CLEAN (MRC), multi-scale Clean and wavelet CLEAN.

\section{Image Deconvolution by GAN}
\label{sect:model}

Recently, deep learning (DL) \cite{goodfellow2014generative,xu2018solar,hinton2006reducing,hinton2006fast,bengio2009learning,lecun1989backpropagation,lecun1998gradient} was intensively developed and achieved big success in many application fields, such as image processing, speech recognition, natural language understanding, pattern recognition and computer vision. The advantages of DL lie in twofold. Firstly, it can learn a model from mass of data, which would be more applicable in practice. While traditional machine learning model or physical model would not make full use of mass of available data. Secondly, DL does not need to fully acquire knowledges of a physical process. It would not establish a mathematical model at all. Instead, an extreme non-learning relation between input and output is learnt in a data-intensive manner. This advantage of DL makes itself possesses more flexibility and applicability.
GAN \cite{goodfellow2014generative,xu2018solar} is a DL model which was recently raised and has been extensively investigated in many kinds of applications, especially image reconstruction, such as image denoising, image synthesis, super-resolution. A GAN is comprised of a generator and a discriminator as shown in Fig. 3. In Fig. 3, both real/original image and fake image are fed into a GAN. The generator tries to make fake images close to real/original ones. Meanwhile, the discriminator acts as a detector to distinguish fake images forged by the generator from real/original ones. By adversarial learning, the generator could acquire the distribution of training samples, so that it can produce new/unknow samples, while the discriminator can tell fakes from originals well. In addition, the discriminator becomes better and better after learning from cheating of the faker (generator) again and again. Repeating this process, finally, the generator could forge data very close to real data, at the same time, the discriminator becomes an excellent detector. The principle of GAN is originated from zero-sum minimax game, which is mathematically represented by
\begin{equation}\label{exp-gan}
 \begin{aligned}
G^* = &\arg\min\limits_G\max\limits_D \mathcal{L}_{GAN}(G,D)\\
\mathcal{L}_{GAN}(G,D)= &\mathbb{E}_y[\log D(y)]+\mathbb{E}_{x, z}[\log (1-D(G(x,z))))]
 \end{aligned}
\end{equation}
where $D$ represents a detector, $G$ represent a generator, $y$ is a real image and $G(x,z)$ is a fake image. In (5), $y$ is coming from a distribution of real data, $x$ is coming from our simulated data (e.g., degraded images in image processing), and $z$ is coming from a random noise. For optimizing $D$, we expect the larger $D(y)$ on the real data and the smaller $D(G(x, z))$ on the fake data generated by the generator G. While for optimizing $G$, we expect that it can generate enough realistic sample $G(x,z)$ to cheat $D$ successfully. During training process, $D$ and $G$ are optimized alternatively, by fixing one and optimizing the other.
\begin{figure}
   \centering
   \includegraphics[width=\textwidth, angle=0]{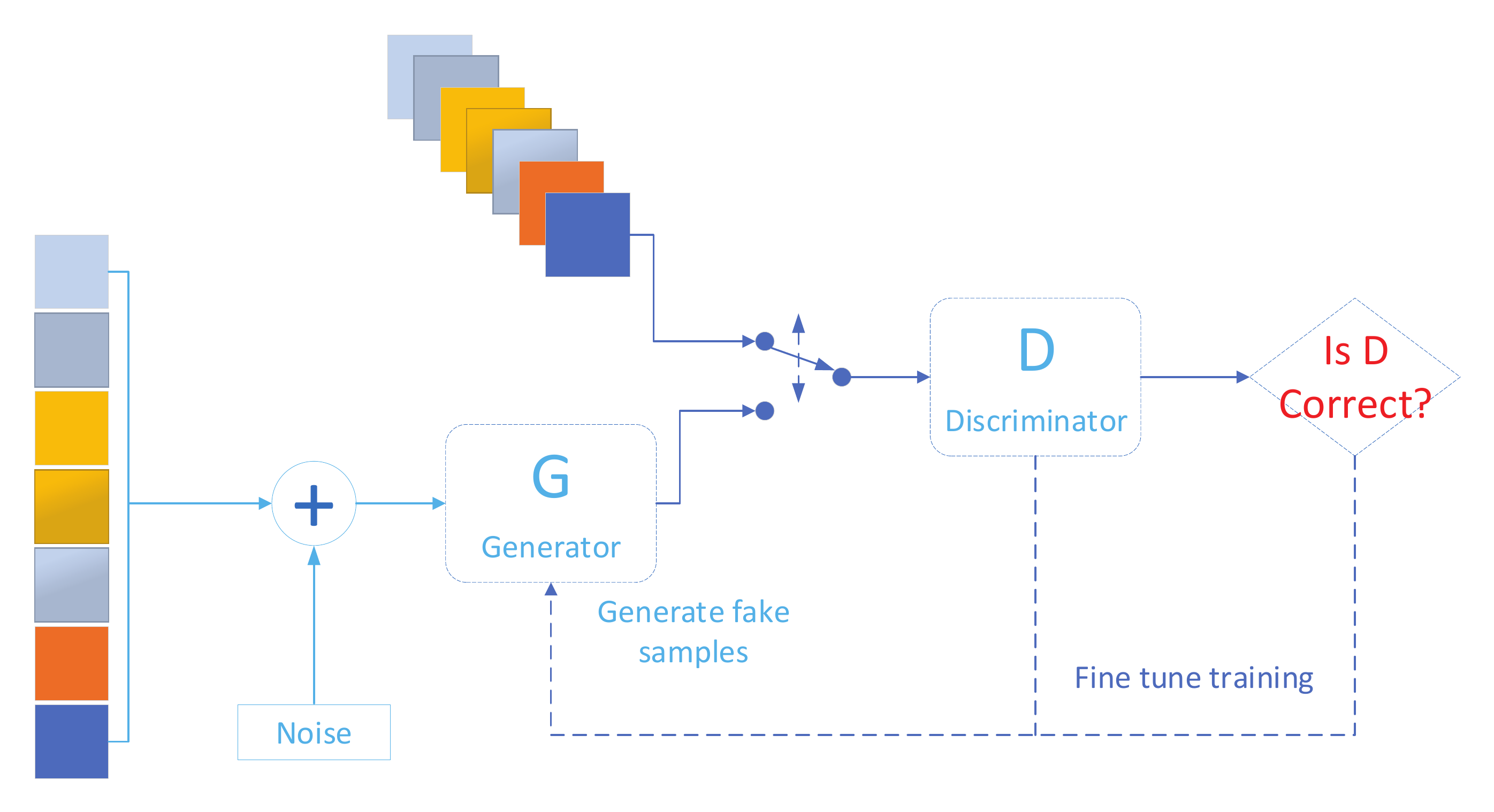}
   \caption{Generative Adversarial Neural Network \cite{ganpicture}.}
   \label{fig:3}
\end{figure}

As (5) indicated, a general GAN only discriminate fake and true of the output. However, most of image processing tasks, e.g., well-known image-to-image translation \cite{2016arXiv161107004I}, require the correspondences between inputs and outputs besides discriminating fake and true. For this purpose, the conditional GAN (cGAN) was proposed, which is described by
\begin{equation}\label{exp-cgan}
 \begin{aligned}
G^* = &\arg\min\limits_G\max\limits_D \mathcal{L}_{GAN}(G,D)\\
\mathcal{L}_{cGAN}(G,D)= &\mathbb{E}_{x,y}[\log D(x,y)]+\mathbb{E}_{x, z}[\log (1-D(G(x,z))))]
 \end{aligned}
\end{equation}
where $D(x,y)$, $D(x, G(x, z))$ indicates that $D$ needs not only distinguish the real and the fake, but also tell the correspondence between them.
In \cite{2016arXiv161107004I}, Phillip Isola et.al. described a cGAN model for image-to-image translation, namely pix2pix. In our case, we refer to pix2pix network for solar image deconvolution, while the optimization objective is revised for facilitating our specific task. Besides cGAN loss and L1 loss of spatial domain($\mathcal{L}_{L1}^{I}(G)=\mathbb{E}_{x,y,z}[\|y-G(x,z)\|_1]$) in pix2pix network \cite{2016arXiv161107004I}, a new loss, namely perceptual loss \cite{Johnson_2016}, is also introduced additionally as,
\begin{equation}\label{lossfun}
\mathcal{L}_{L1}^{P}(G)=\mathbb{E}_{x,y,z}[\|\Phi(y)-\Phi(G(x,z))\|_1]
\end{equation}
where $\Phi(\cdot)$ represents the feature of an image, specifically, VGG feature from a pre-trained VGG-16 model \cite{Simonyan2014Very}. Here, the feature maps of the first four layers of a VGG-16 network are extract to give $\Phi(y)$ and $\Phi(G(x,z))$. Thus, the final objective is
\begin{equation}
G^*=\arg\min\limits_G\max\limits_D \mathcal{L}_{cGAN}(G,D) + \lambda_1\mathcal{L}_{L1}^{I}(G) + \lambda_2\mathcal{L}_{L1}^{P}(G)
\end{equation}

In our model, the generator is a classical UNet, consisting of multiple layers of convolution and transposed convolution as illustrated in Fig. 4. From Fig. 4, UNet is in the shape of an auto-encoder. The encoder gets compressed representation of the input, while the decoder decompresses this representation to reconstruct the input. The most noteworthy feature of UNet is the skip connection between corresponding layers of the encoder and the decoder. This skip connection can combine both high level semantic information and low level features of an image, benefiting image processing tasks, especially for images with less semantic information, like medical and astronomy images. The discriminator is a general convolution neural network consisting of 5 convolution layers.

\begin{figure}
   \centering
   \includegraphics[width=\textwidth, angle=0]{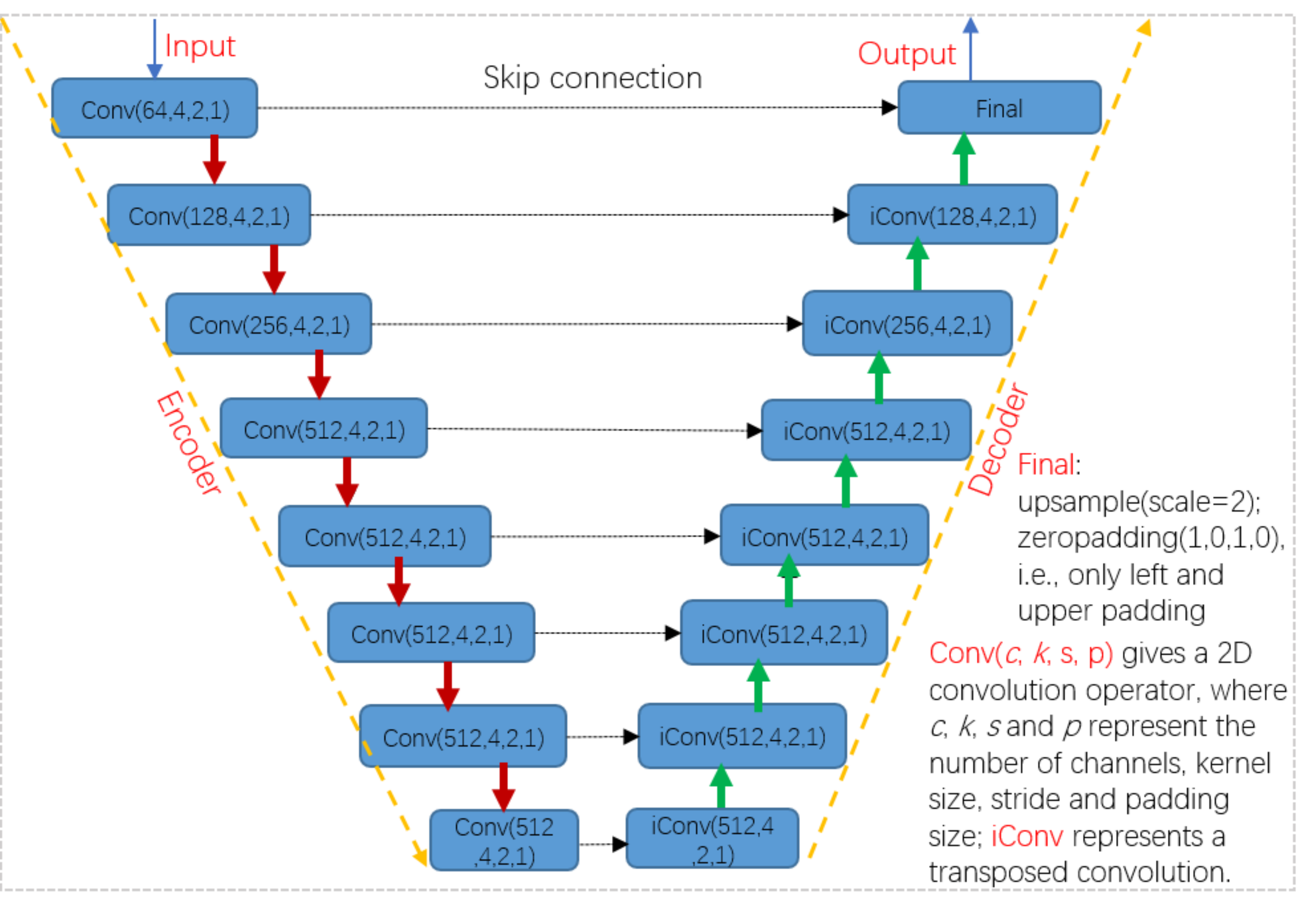}
   \caption{The proposed model for AS image deconvolution. }
   \label{fig:4}
\end{figure}
Image generation/reconstruction, like image deblurring, denoising and super-resolution, has been well investigated in the literature \cite{2017arXiv171107064K,DeblurGAN,2016arXiv161202177N,yan2017dcgans}. Image deconvolution is a typical image generation problem. Usually, in radio astronomy, it was handled by ``CLEAN" algorithm \cite{hogbom1974aperture,wakker1988multi,cornwell2008multiscale}. Two conditions should be held for the success of the CLEAN algorithm on image deconvolution. One is that the signal should be point source, the other is that dirty beam should be exactly known, which means the dirty beam of actual system and the ideal one are exactly the same. However, in practice, these two conditions do not hold so that the efficiency of the CLEAN algorithm is compromised. The proposed model is learnt from data without any constraint, which is completely data-driven, so it has more competitive advantages in the era of big data.

\section{Experimental results}
\label{sect:model}
To evaluate the proposed model, a database consisting original/clear and dirty image pairs is firstly established. We collected 41, 096 images of 193 {\AA} from Atmospheric Imaging Assembly (AIA) of Solar Dynamics Observatory (SDO) as ground-truth/clear images. Then, we apply MUSER-I dirty beam (as shown in Fig. 1(b)) to these clear images, resulting in corresponding dirty images. For training, validation and testing, the database is split into 3 parts: 8000 image pairs for validation, 8000 image pairs for testing. The full implementation (based on Pytorch) and the trained network can be accessed via \url{https://github.com/lowenve/solarGAN}. From the statistics of experimental results, we can observe and conclude that:
\begin{enumerate}[1)]
\item
In the beginning, the generated image is with low quality since the training process is far from convergence;
\item
After about 5000 loops, the learnt model can be stable, generating high quality images as shown in Fig. 5, where the left column gives dirty images, the middle column shows output images after GAN deconvolution, the right column shows original images;
\item
The learnt model can restore image details/structures well, as shown in Fig. 5(b). Compared with dirty image in Fig.5 (a), the reconstructed one contains more details of an image;
\item
We also verify the effectiveness of spatial loss and perceptual loss as claimed in [\ref{lossfun}] for our task. The PSNR and SSIM on the whole testing dataset are compared in Table \ref{exp-loss}. It can be observed that the best result is coming from the combination of cGAN loss, spatial-domain L1 loss and perceptual L1 loss.
\end{enumerate}

\begin{table}
\footnotesize
\begin{center}
\caption{Performance verification of the proposed network with different loss function.}
\label{exp-loss}
\begin{tabular}{c|c|c}
\hline
   Loss function & PSNR & SSIM\\
\hline
$\mathcal{L}_{{cGAN}(G,D)}+\mathcal{L}_{L1}^{I}$	&38.0575	&0.9561	\\
\hline
$\mathcal{L}_{{cGAN}(G,D)}+\mathcal{L}_{L1}^{I}+\mathcal{L}_{L1}^{P}$	&$\mathbf{38.4442}	$ & $\mathbf{0.9609}$\\
\hline
$\mathcal{L}_{{cGAN}(G,D)}+\mathcal{L}_{L2}^{I}$	&35.2378	&0.9316\\
\hline
$\mathcal{L}_{{cGAN}(G,D)}+\mathcal{L}_{L2}^{I}+\mathcal{L}_{L2}^{P}$	&37.3543	&0.94727\\
\hline
\end{tabular}
\end{center}
\end{table}

For objective measurement of image quality, peak signal to noise ratio (PSNR) and structural similarity index measurement (SSIM) \cite{wang2004image} are employed for evaluating the proposed model. PSNR measures the absolute difference of pixel-to-pixel of two images. SSIM may ignore the pixel-to-pixel difference, while pays more attention to the similarity of image structure. The PSNR and SSIM statistics are listed in Table \ref{Tab:6}. From Table \ref{Tab:6}, the remarkable improvements of image quality can be achieved by the proposed model, where the average PSNR improvement of 5dB and average SSIM improvement of 5.5\% are achieved by the proposed GAN-based model.

In real situation, small disturbance exists, which will make the situation more complicated, so the model should be more flexible and robust for addressing these complicated situations. For this purpose, we introduce one of the noises, namely, Gaussian white noise, in our simulation for checking the flexibility and robustness of the proposed model. The experiment results, as listed in Table \ref{Tab:6}, prove that the proposed model still applicable and keep the same efficiency of the situation without noise. From Table \ref{Tab:6}, the PSNR improvement achieves 4.11dB. Remarkably, the SSIM improvement is up to 16\%. Further analysis reveals that the SSIM is sensitive to Gaussian white noise, so image quality measured by SSIM drops a lot after contaminated by Gaussian white noise. Fortunately, the proposed model could learn the abilities of not only deconvolution but also denoising, so the SSIM improvement is remarkable. The conclusion is that the proposed model is robust for handling small noise, partially because it is a data driven model.

\begin{table}
\footnotesize
\begin{center}
\caption{Performance comparisons.}
\label{Tab:6}
\centering
\begin{tabular}{c|c|c|c|c|c|c}
\hline
\multirow{2}{*}{Test images} & \multicolumn{2}{|c|}{PSNR(dB)} & \multicolumn{2}{|c|}{SSIM(block:8x8)} & \multicolumn{2}{|c|}{SSIM(block:16x16)}\\
\cline{2-7}
   & deconvolved & dirty& deconvolved & dirty& deconvolved & dirty\\
\hline
2012-08-31 19:48:06UT	&43.4474	&38.1938	&0.9512	&0.8965	&0.9493	&0.8932\\
\hline
2014-09-17 08:48:06UT	&43.7925	&38.3835	&0.9527	&0.9004	&0.9509	&0.8971\\
\hline
2014-09-17 09:00:06UT	&43.7546	&38.2539	&0.9520	&0.8975	&0.9502	&0.8942\\
\hline
2014-09-17 09:12:06UT	&42.9609	&37.6210	&0.9508	&0.8887	&0.9491	&0.8847\\
\hline
2015-05-28 12:48:06UT	&43.3054	&38.5841	&0.9530	&0.9028	&0.9512	&0.8988\\
\hline
2016-05-18 02:00:05UT	&43.2166	&38.5393	&0.9531	&0.9020	&0.9512	&0.8981\\
\hline
2016-05-18 02:12:05UT	&43.2083	&38.5592	&0.9504	&0.9014	&0.9484	&0.8979\\
\hline
2017-02-01 03:48:04UT	&43.1904	&38.4375	&0.9500	&0.8987	&0.9481	&0.8951\\
\hline
2017-02-01 04:00:04UT	&44.0610	&39.4934	&0.9574	&0.9174	&0.9558	&0.9140\\
\hline
2017-09-03 00:48:04UT	&44.8305	&40.0187	&0.9599	&0.9252	&0.9584	&0.9223\\
\hline
$\mathbf{Average}$ 	&$\mathbf{43.5768}$	&$\mathbf{38.6084}$	&$\mathbf{0.9531}$	&$\mathbf{0.9031}$	&$\mathbf{0.9513}$	&$\mathbf{0.8995}$\\
\hline
$\mathbf{Improvement}$ & \multicolumn{2}{|c|}{$\mathbf{4.9683}$}  &\multicolumn{2}{|c|}{$\mathbf{0.0500}$} &\multicolumn{2}{|c|}{$\mathbf{0.0517}$}\\
\hline\hline
2012-08-31 19:48:06UT   &41.4628   &36.8767    &0.9456    &0.8076    &0.9435    &0.8035\\
   \hline
2014-09-17 08:48:06UT   &41.6365   &37.0072    &0.9474    &0.8107    &0.9454    &0.8067\\
   \hline
2014-09-17 09:00:06UT   &41.5098   &36.9217    &0.9457    &0.8086    &0.9437    &0.8046\\
   \hline
2014-09-17 09:12:06UT   &41.0332   &36.4721    &0.9424    &0.7998    &0.9404    &0.7955\\
   \hline
2015-05-28 12:48:06UT   &41.3615   &37.1773    &0.9489    &0.8156    &0.9469    &0.8116\\
   \hline
2016-05-18 02:00:05UT   &41.0803   &37.1395    &0.9487    &0.8146    &0.9467    &0.8106\\
   \hline
2016-05-18 02:12:05UT   &40.3568   &37.1832    &0.9467    &0.8141    &0.9446    &0.8101\\
   \hline
2017-02-01 03:48:04UT   &40.0558   &37.0995    &0.9451    &0.8120    &0.9430    &0.8079\\
   \hline
2017-02-01 04:00:04UT   &41.6432   &37.7621    &0.9547    &0.8272    &0.9529    &0.8238\\
   \hline
2017-09-03 00:48:04UT	&42.7742   &38.0829    &0.9580    &0.8341    &0.9564    &0.8309\\
   \hline
$\mathbf{Average}$ 	&$\mathbf{41.2914}$	&$\mathbf{37.1722}$	&$\mathbf{0.9483}$	&$\mathbf{0.8144}$	&$\mathbf{0.9464}$	&$\mathbf{0.8105}$\\
\hline
$\mathbf{Improvement}$ & \multicolumn{2}{|c|}{$\mathbf{4.1192}$}  &\multicolumn{2}{|c|}{$\mathbf{0.1339}$} &\multicolumn{2}{|c|}{$\mathbf{0.1358}$}\\
\hline
\end{tabular}
\end{center}
\end{table}

For comparison between the proposed model and traditional H\"{o}gbom CLEAN, the dirty image in Fig. 5(a) is processed by H\"{o}gbom CLEAN. The results of H\"{o}gbom CLEAN are shown in Fig. 6, where Figs. 6(a) and (b) demonstrate the images of bright points after H\"{o}gbom CLEAN of 400 and 4000 iterations, respectively. Fig. 6(c) is the residual image corresponding to Fig. 6(b). Fig. 6(d) gives the final deconvoluted image which combines the residual image (Fig. 6(c)) and the image of bight points (Fig. 6(b)). From Fig. 6, it can be concluded that H\"{o}gbom CLEAN can successfully restore bright points in an image, however fail to restore image details. This conclusion also confirms that H\"{o}gbom CLEAN is designed for point source instead of extended source. Comparing Fig. 6 and Fig. 5, the proposed model is dramatically superior to H\"{o}gbom CLEAN on restoring image details/fine structures.

   \begin{figure}
   \begin{minipage}{1.0\textwidth}
   \includegraphics[width=0.33\textwidth, angle=0]{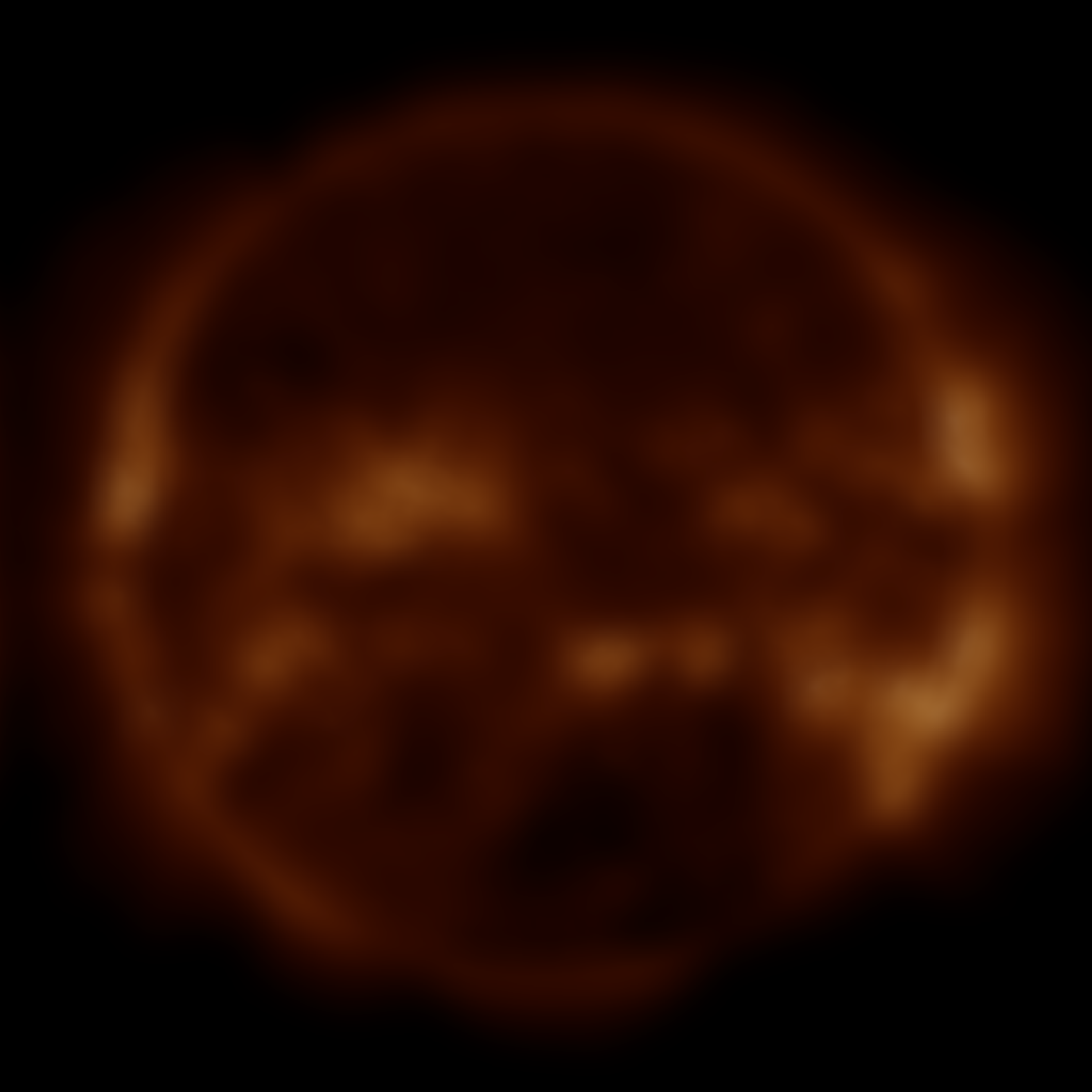}
   \includegraphics[width=0.33\textwidth, angle=0]{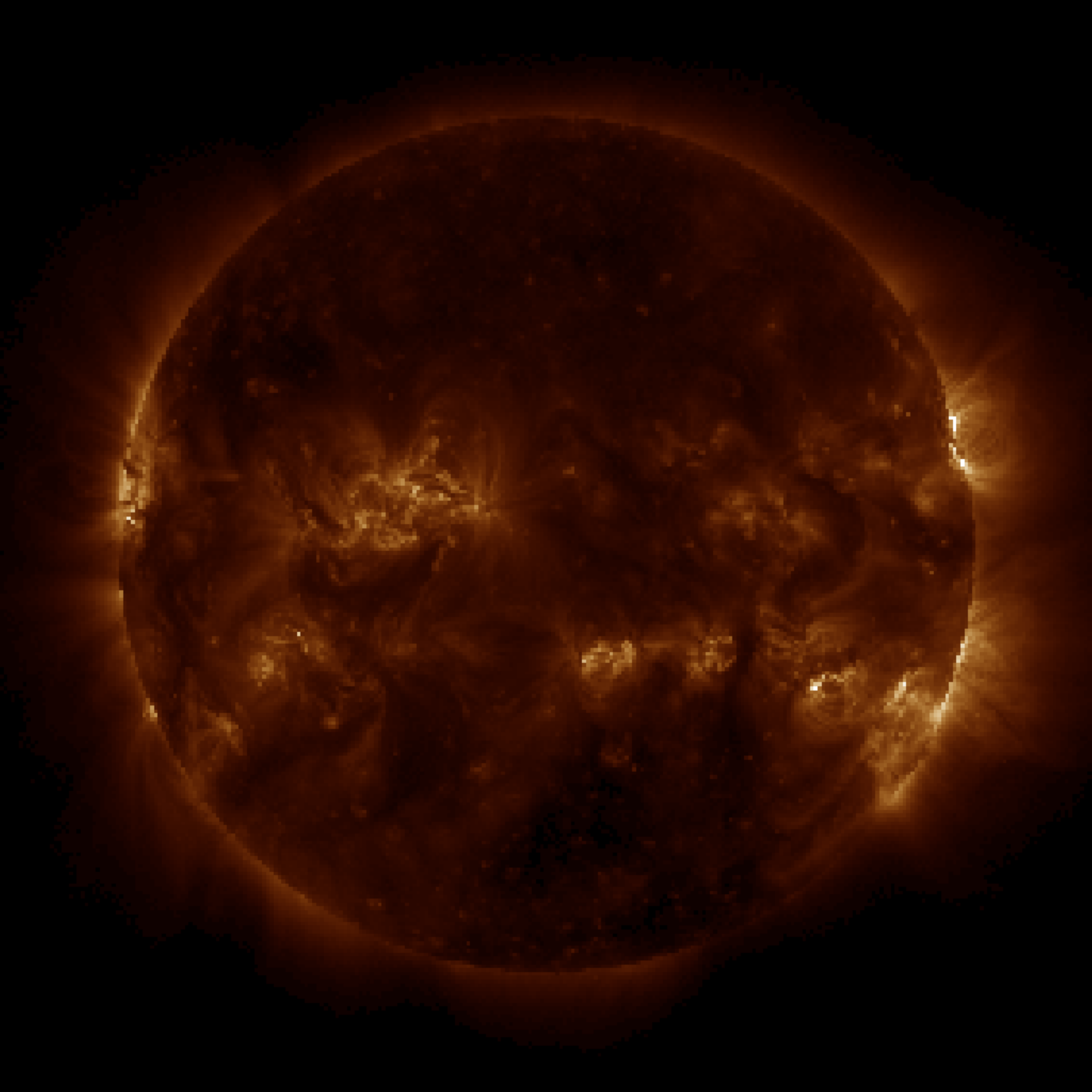}
   \includegraphics[width=0.33\textwidth, angle=0]{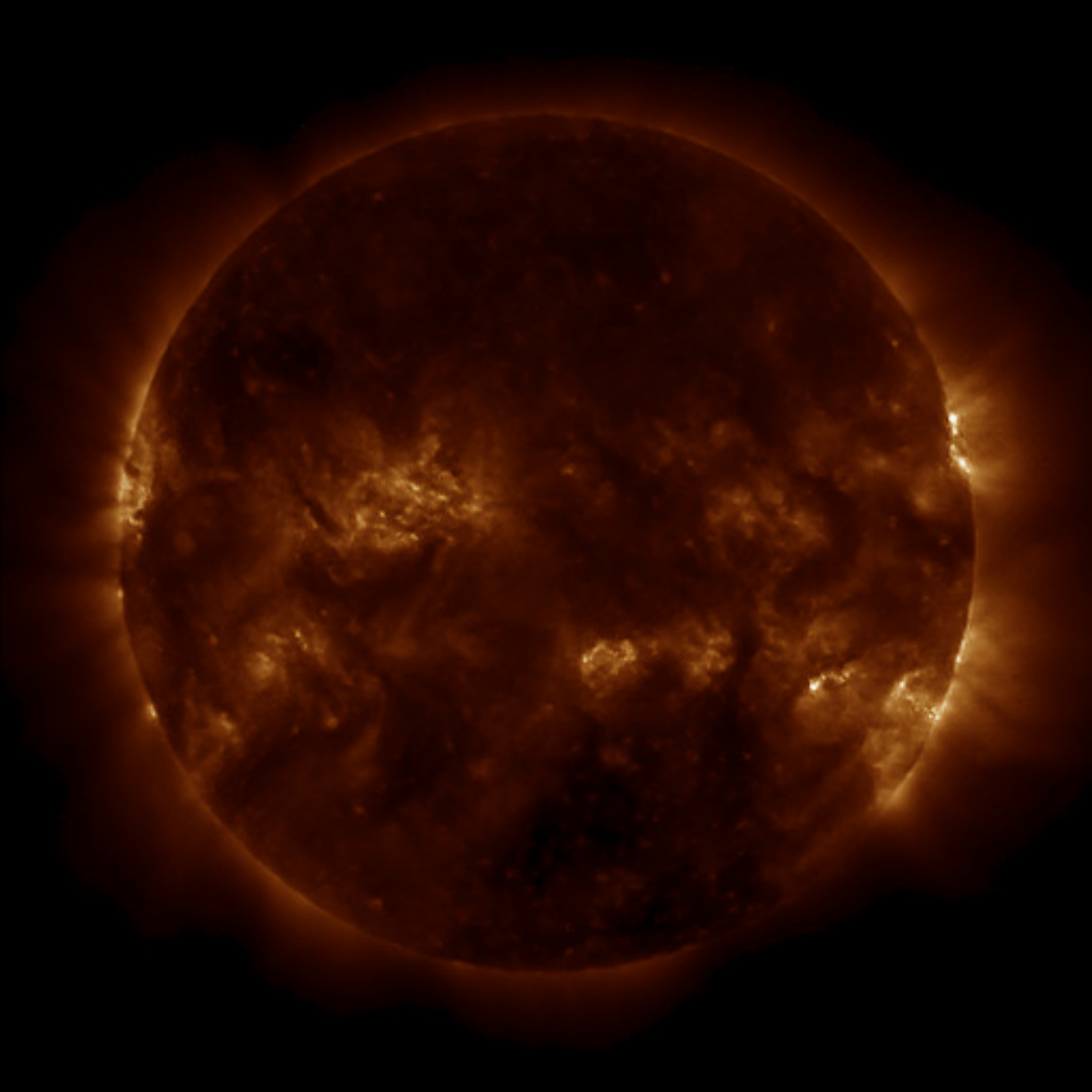}
   \centerline{(a) 2014-09-17, 09:00 (from left to right: dirty, deconvolved and original images)}
   \end{minipage}
   \begin{minipage}{1.0\textwidth}
   \includegraphics[width=0.33\textwidth, angle=0]{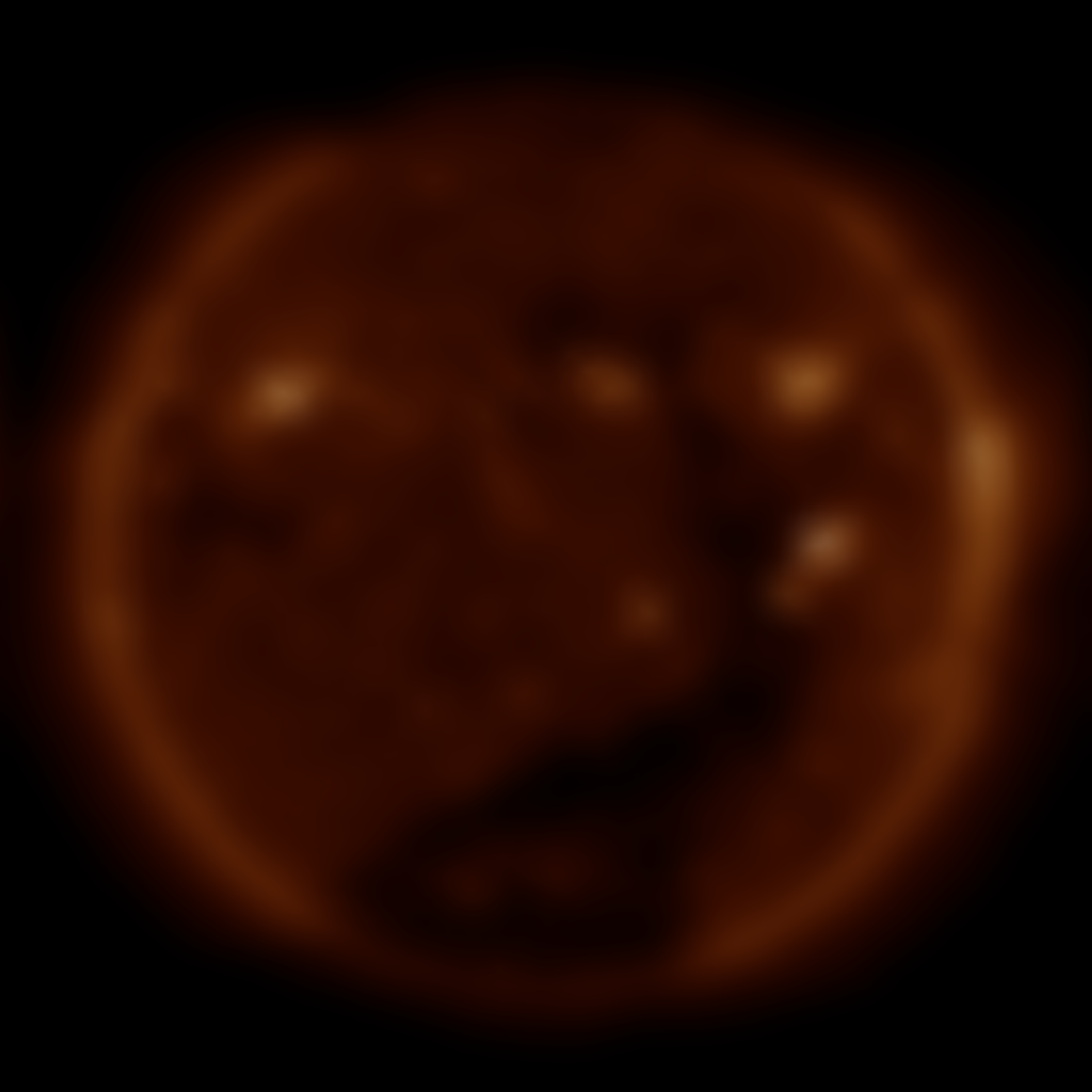}
   \includegraphics[width=0.33\textwidth, angle=0]{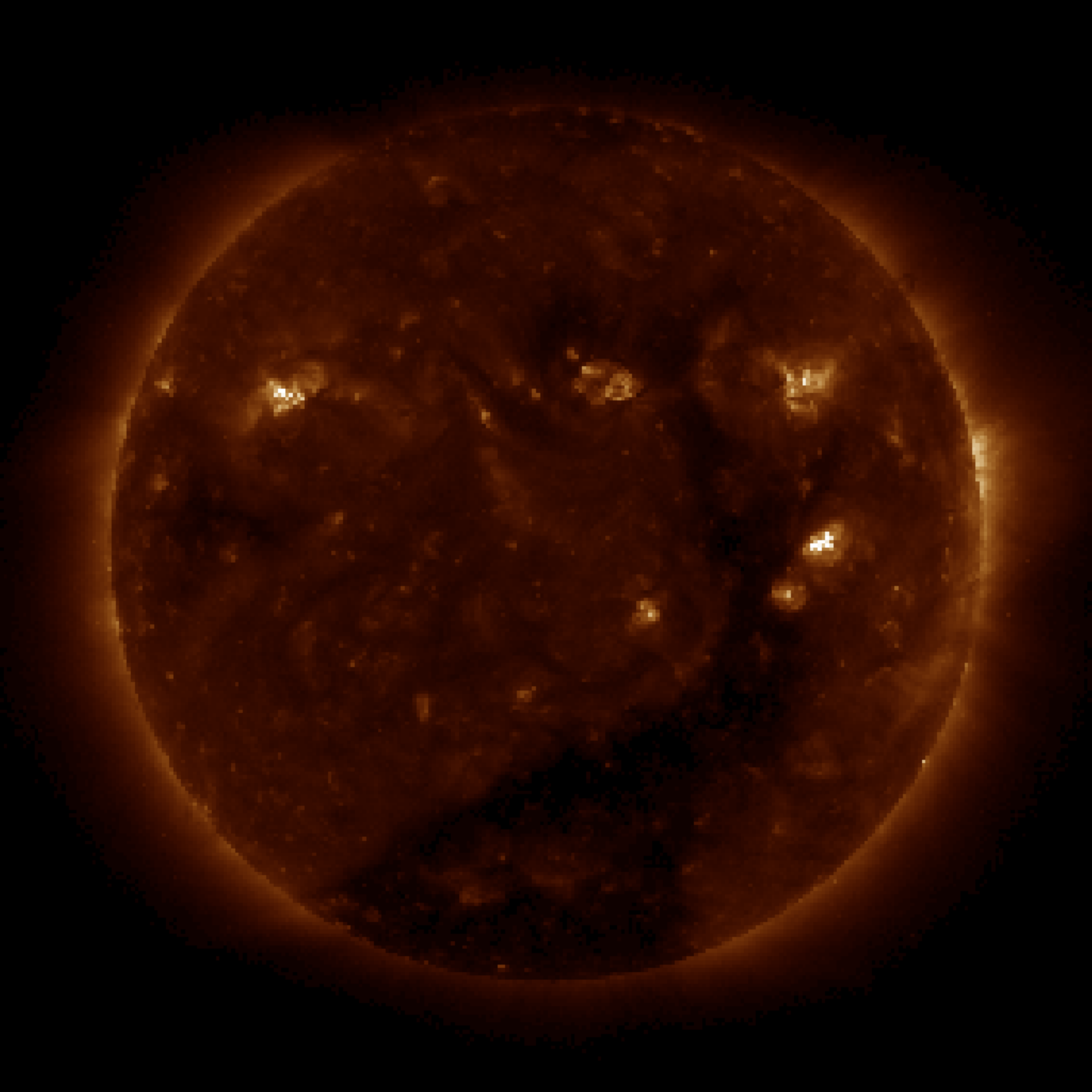}
   \includegraphics[width=0.33\textwidth, angle=0]{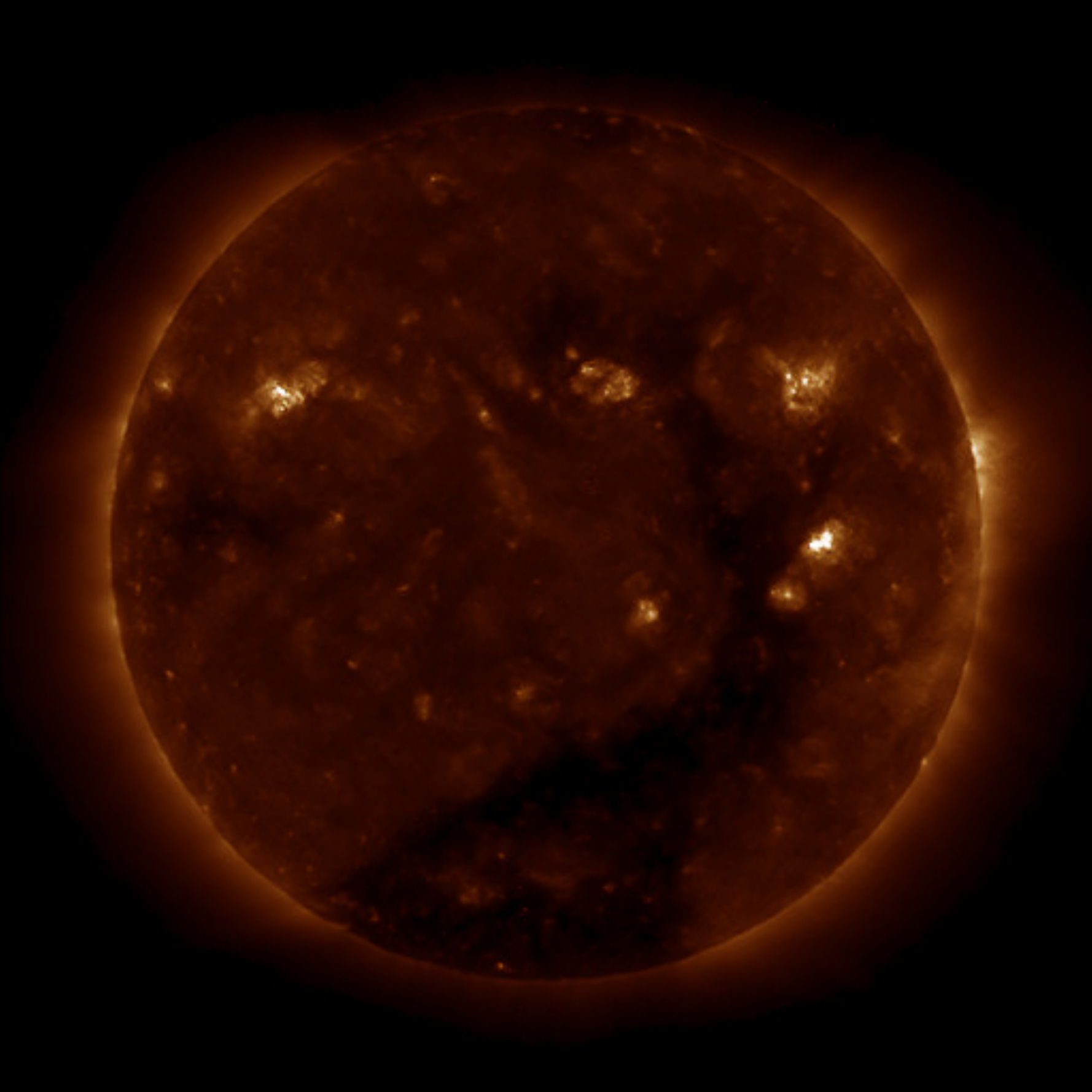}
   \centerline{(b) 2017-02-01, 03:48 (from left to right: dirty, deconvolved and original images)}
   \end{minipage}
   \begin{minipage}{1.0\textwidth}
   \includegraphics[width=0.33\textwidth, angle=0]{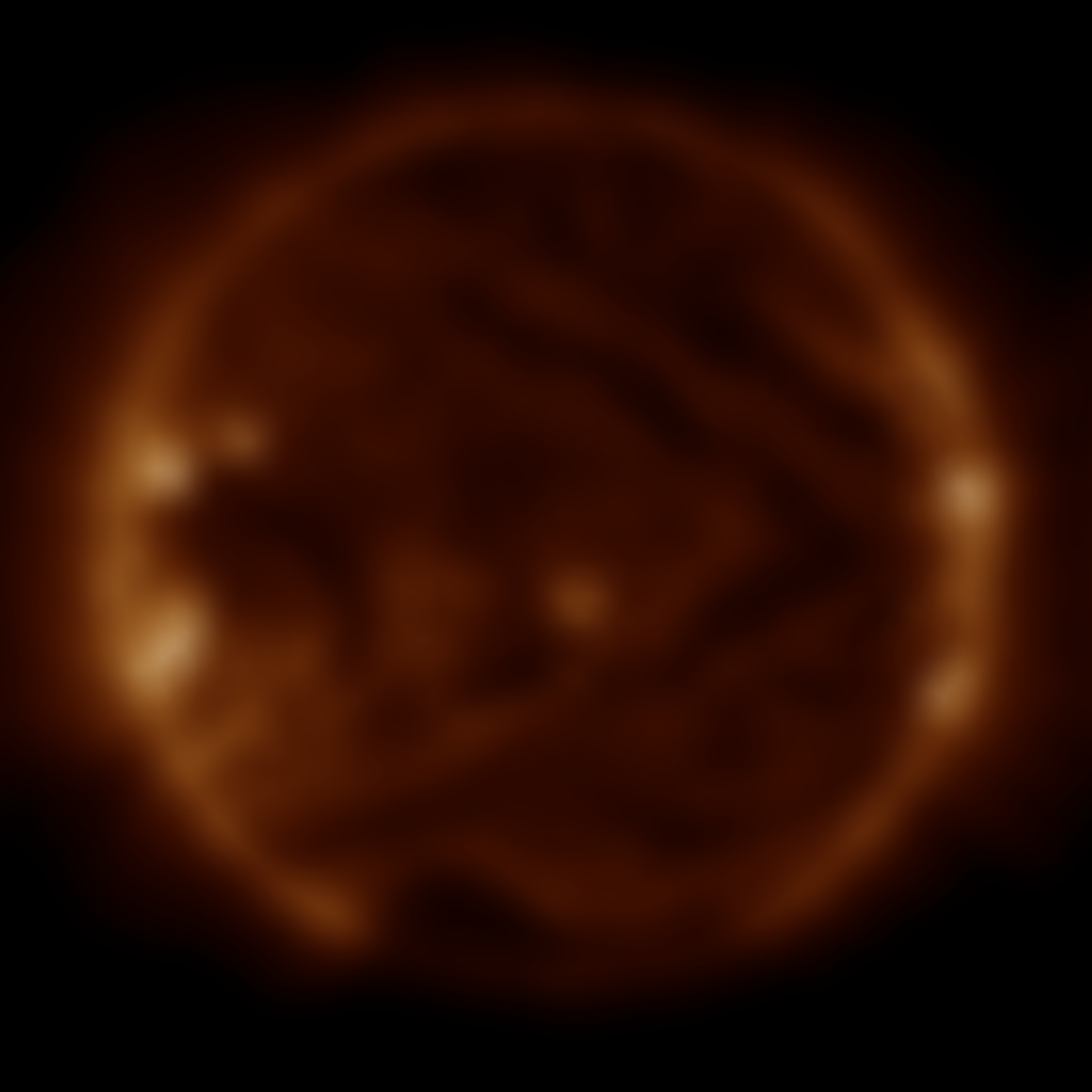}
   \includegraphics[width=0.33\textwidth, angle=0]{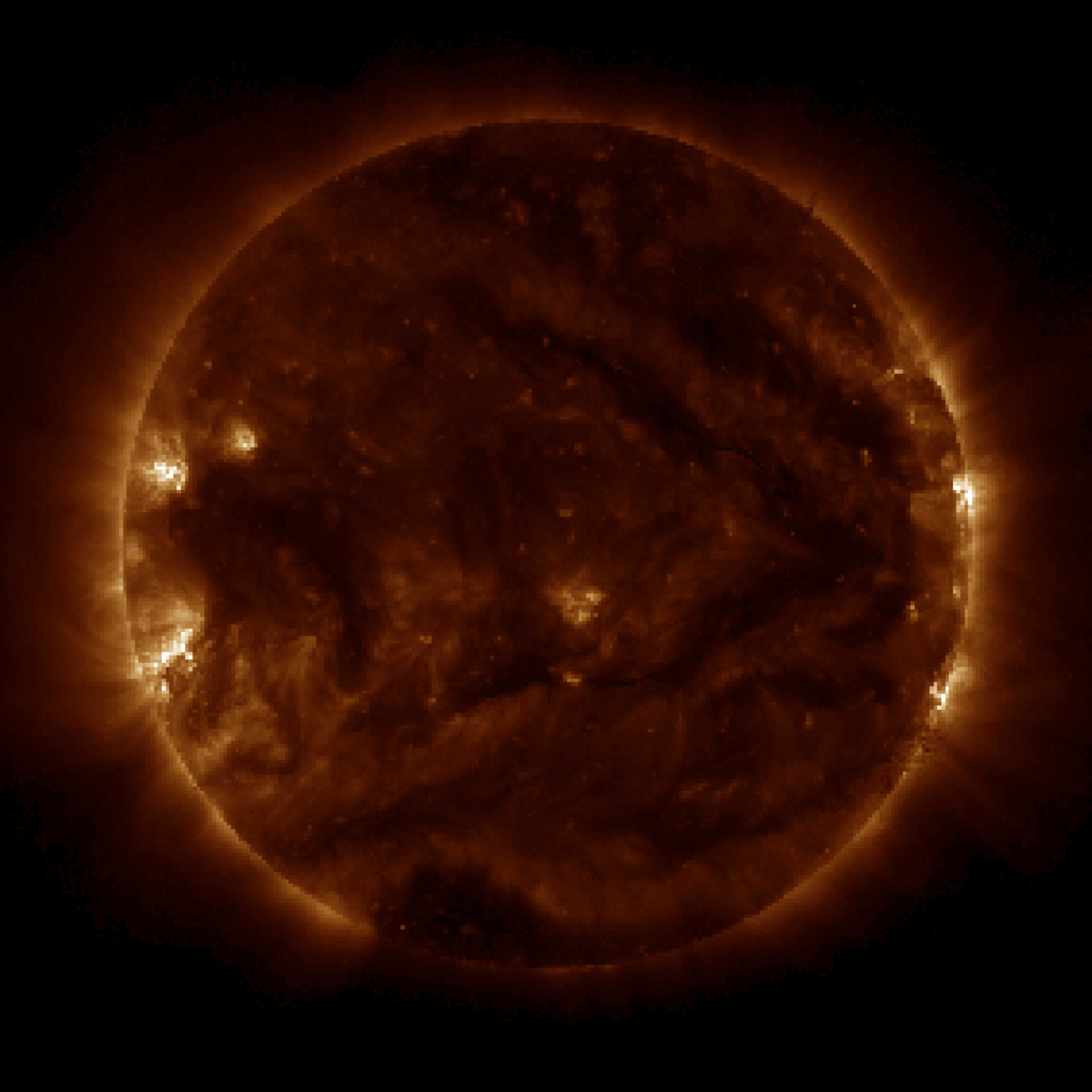}
   \includegraphics[width=0.33\textwidth, angle=0]{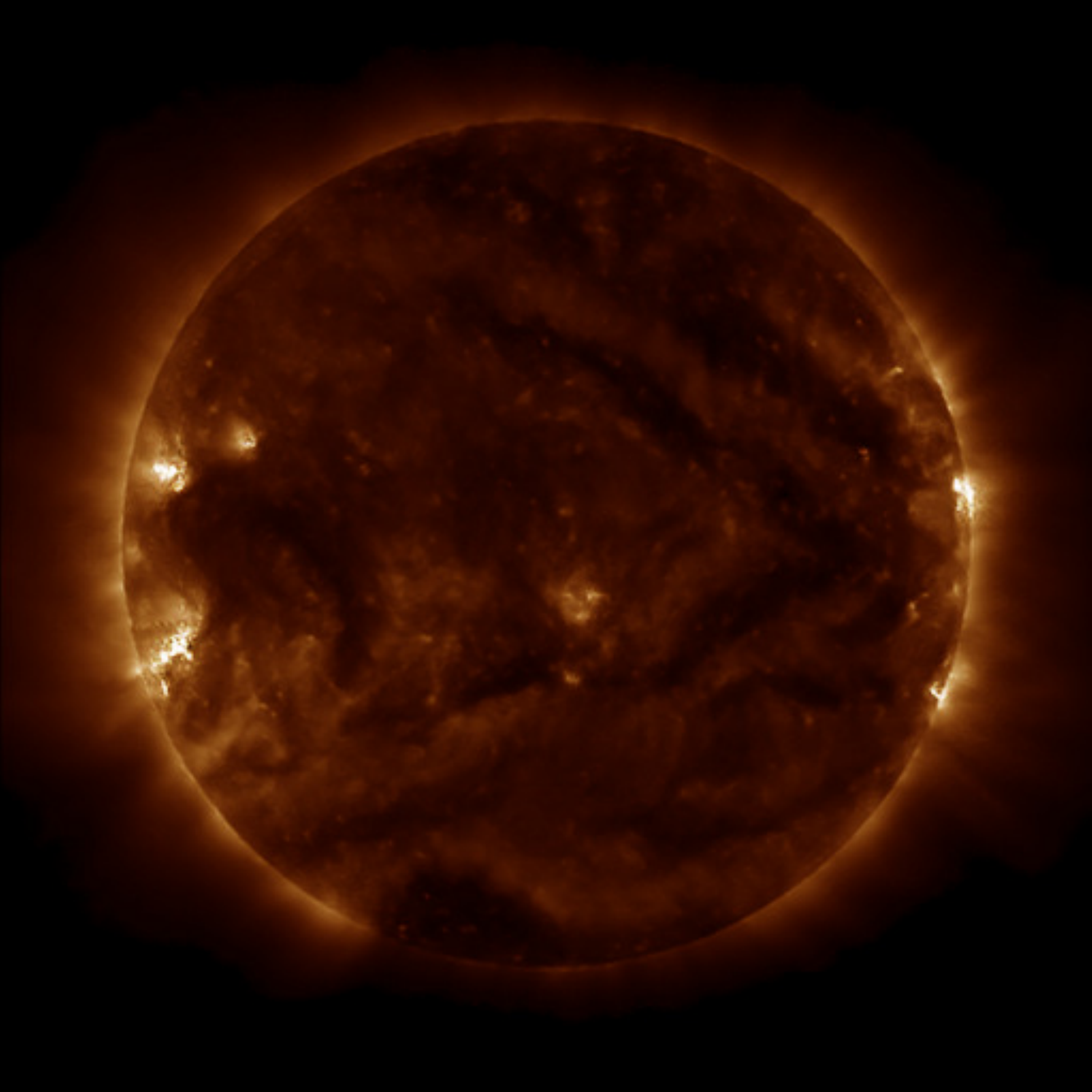}
   \centerline{(c) 2014-09-17, 09:12 (from left to right: dirty, deconvolved and original images)}
   \end{minipage}
   \caption{Image quality comparison between dirty images (left), deconvolved images (middle) and original one (right) (SDO/AIA, 193 {\AA}, dirty images are derived from MUSER-I sampling)}
   \label{fig:5}
   \end{figure}

   \begin{figure}
   \begin{minipage}[t]{0.24\linewidth}
   \centering
   \includegraphics[width=1.0\textwidth, height=1.0\textwidth,angle=0]{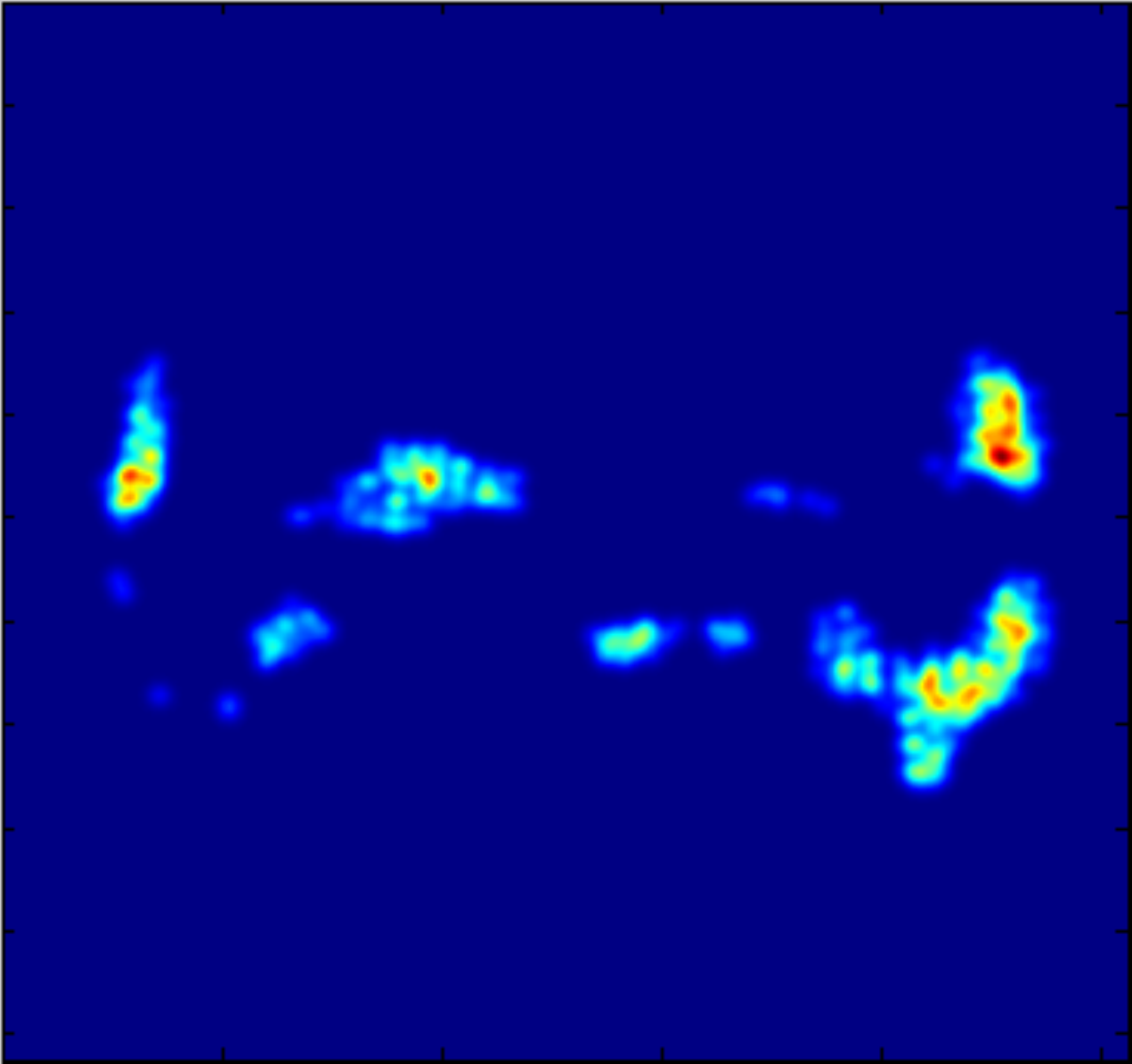}
   \caption*\scriptsize{{(a) Bright points of 400 iterations}}
   \end{minipage}
   \label{fig6a}
   \begin{minipage}[t]{0.24\linewidth}
   \centering
   \includegraphics[width=1.0\textwidth, height=1.0\textwidth,angle=0]{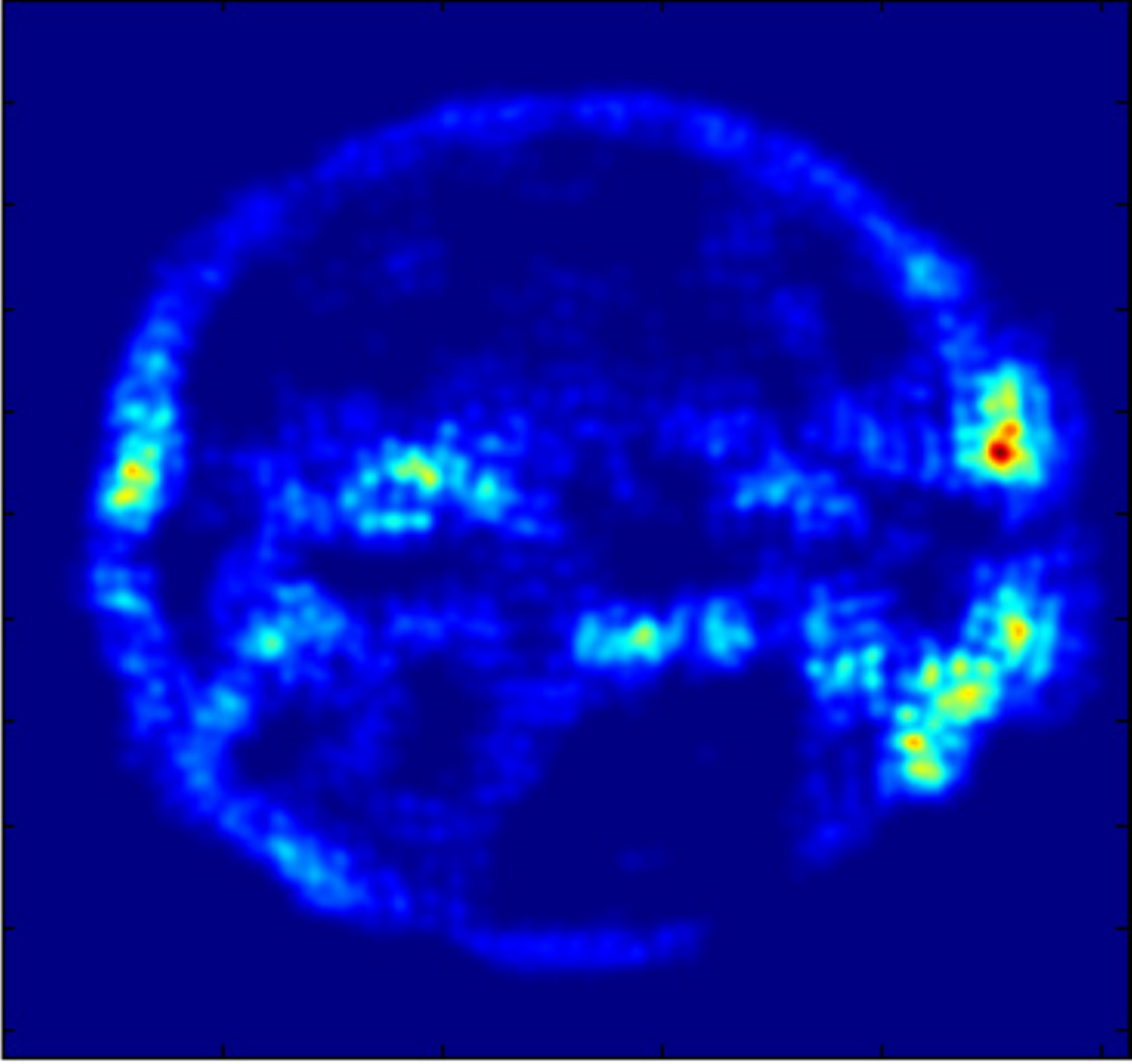}
   \caption*\scriptsize{{(b) Bright points of 4000 iterations}}
   \end{minipage}
   \label{fig6b}
   \begin{minipage}[t]{0.24\linewidth}
   \centering
   \includegraphics[width=1.0\textwidth, height=1.0\textwidth,angle=0]{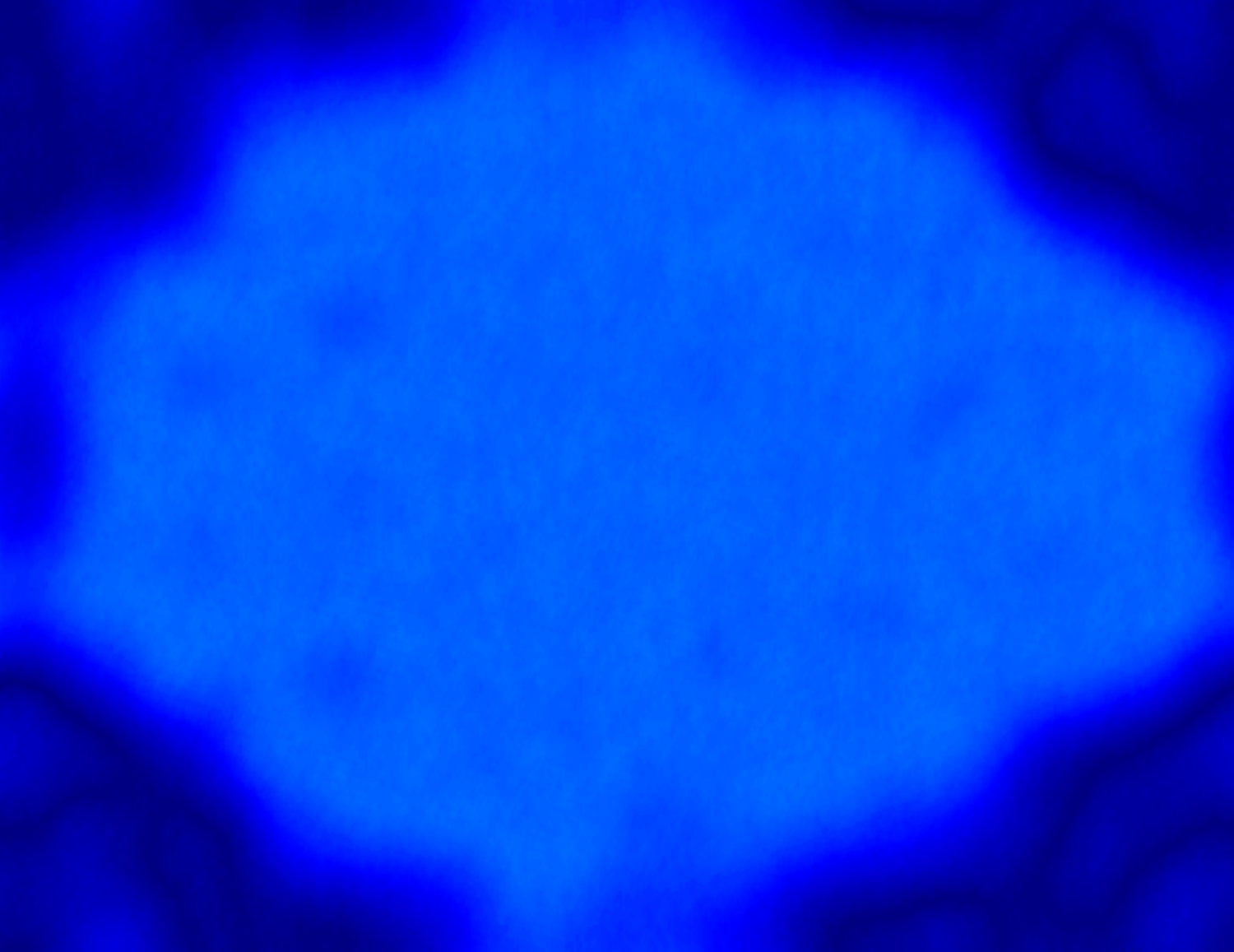}
   \caption*\scriptsize{{(c) Residual image after 4000 iterations}}
   \end{minipage}
   \label{fig6c}
   \begin{minipage}[t]{0.24\linewidth}
   \centering
   \includegraphics[width=1.0\textwidth, height=1.0\textwidth,angle=0]{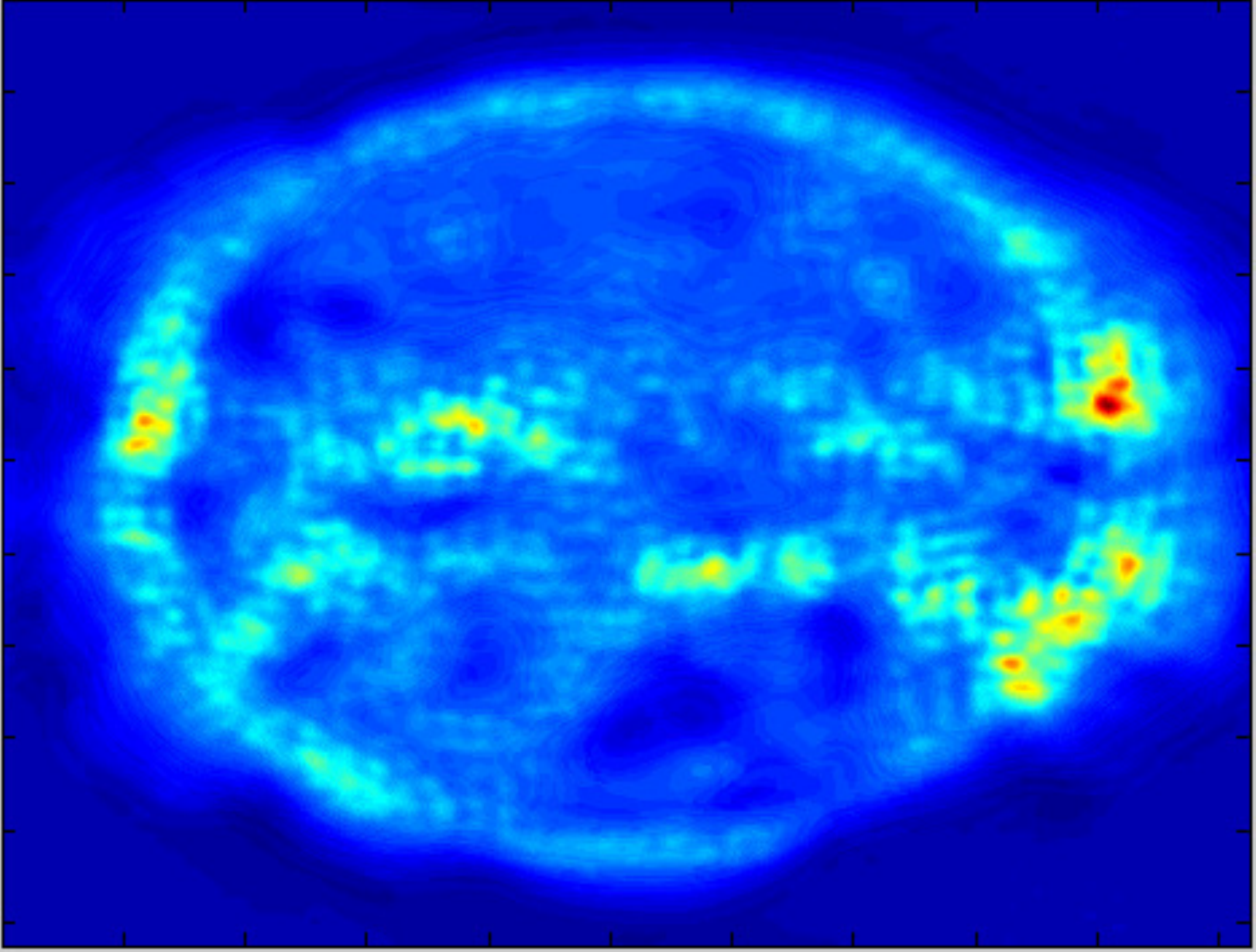}
   \caption*\scriptsize{{(c) Final deconvoluted image by H\"{o}gbom CLEAN}}
   \end{minipage}
   \label{fig6d}
   \caption{The reconstructed image by using H\"{o}gbom CLEAN ((a) and (b) only shows bright points without quiet solar background; here only grayscale images are processed since H\"{o}gbom CLEAN is implemented on grayscale image)}
   \label{fig:6}
   \end{figure}

\section{Conclusions}
\label{sect:conclusion}
This paper makes an effort on a deep learning model for image deconvolution. Given original and dirty image pairs, a generator can be trained from the GAN framework. This generator can then be used to accomplish deconvolution. The evaluations on solar image demonstrate that the proposed model can recover image details markedly better than the traditional CLEAN, so it is fit for extended source, like the sun. In addition, our model is data-driven instead of physical model. For this reason, it is more applicable under some complex situations, even unknown the PSF.

\begin{acknowledgements}
This work was partially supported by a grant from the National Natural Science Foundation of China under Grant 61572461, 11790305, 61811530282, 61872429, 61661146005, U1611461 and CAS 100-Talents (Dr. Xu Long).
\end{acknowledgements}




\bibliographystyle{raa}
\bibliography{RAAsubmission}


\label{lastpage}

\end{document}